\documentclass[aps,prb,twocolumn,superscriptaddress,longbibliography]{revtex4-2}
\usepackage[colorlinks=true,linkcolor=blue,citecolor=blue,urlcolor=blue]{hyperref}
\usepackage{amsmath,amssymb}
\usepackage[utf8]{inputenc}
\usepackage[T1]{fontenc}
\usepackage{url}
\usepackage{adjustbox}
\usepackage{subcaption}
\usepackage{graphicx}
\usepackage{bm}
\usepackage{bbm}
\usepackage{xcolor}
\usepackage{color}
\usepackage{ulem}
\usepackage{subfloat}
\usepackage{ragged2e}
\usepackage{caption}
\usepackage{braket}
\DeclareCaptionFormat{myformat}{\justifying#1#2#3}
\newcommand\orcidID[1]{\href{https://orcid.org/#1}{\orcidicon}} 

\newcommand{\Blue}[1]{{\textcolor{blue}{#1}}}

\newcommand{\be}{\begin {equation}}
\newcommand{\ee}{\end {equation}}
\newcommand{\bs}{\boldsymbol}

\newcommand{ \refb}[1]{\textcolor{blue}{(\ref{#1})}}

\begin{document}
 
\title{Quasi-Majorana modes in the $p$-wave Kitaev chains on a square lattice}

\author{S. Srinidhi}
\email{s.srinidhi321@gmail.com}
\affiliation{Department of Physics, Birla Institute of Technology and Science-Pilani, Pilani, Rajasthan 333031, India}

\author{Aayushi Agrawal}
\email{aagrawal@kias.re.kr}
\affiliation{School of Physics, Korea Institute for Advanced Study, Seoul 02455, Republic of Korea}

\author{Jayendra N. Bandyopadhyay}
\email{jnbandyo@gmail.com}
\affiliation{Department of Physics, Birla Institute of Technology and Science-Pilani, Pilani, Rajasthan 333031, India}

\begin{abstract} 

The topological characteristics of the $p$-wave Kitaev chains on a square lattice with nearest-neighbor and next-nearest-neighbor inter-chains hopping and pairing are investigated. Besides gapless exact zero-energy modes, this model exhibits topological gapless phase hosting edge modes, which do not reside strictly at zero energy. However, these modes can be distinguished from the bulk states. These states are known as pseudo- or quasi-Majorana Modes (qMMs). The exploration of this system's bulk spectrum and Berry curvature reveals singularities and flux-carrying vortices within its Brillouin zone. These vortices indicate the presence of four-fold Dirac points arising from two-fold degenerate bands. Examining the Hamiltonian under a cylindrical geometry uncovers the edge properties, demonstrating the existence of topological edge modes. These modes are a direct topological consequence of the Dirac semimetal characteristics of the system. The system is analyzed under open boundary conditions to distinguish the multiple MZMs and qMMs. This analysis includes a study of the normalized site-dependent local density of states, which pinpoints the presence of localized edge states. Additionally, numerical evidence confirms the robustness of the edge modes against disorder perturbations. The emergence of topological edge states and Dirac points with zero Chern number indicates that this model is a weak topological superconductor.
\end{abstract}

\maketitle

\section{Introduction}

Ongoing research has actively explored various topological materials in the last few years due to their potential technological applications \cite{Yan_2012,Bernevig2022-zx,Yan2017-cp,Wieder2021-rr,Culcer_2020}. It also provides fertile ground for discovering quasi-particles and phenomena predicted in high-energy physics. The study of superconductivity in topological materials, including Dirac \cite{3D_Dirac,2D_Dirac,Exp_3_DSM,CA_DSM} and Weyl semimetals \cite{3DW_DSM,Exp_weyl,WeylSSH2,WeylSSH}, has paved the way to the rapidly evolving field of research in topological condensed matter. This study primarily focuses on the role of a superconducting phase characterized by certain topological material characteristics. Topological superconductors (TSC) have nodal superconducting gaps that give rise to numbers called a topological charge or invariants \cite{Sato_2017,Bernevig2013,Leijnse_2012,shaina24}. The presence of the nodal superconducting gaps defines the variety of TSCs in general. A characteristic feature of TSCs is the existence of gapless boundary states that support the existence of degenerate states, like the topological defects in the system. A defect can be viewed as a hole (or a singularity) in a closed contour that leads to the topological zero-energy modes in the system if they are localized in a vortex core. These zero-energy states are often localized at the edge of the material and are robust against local perturbations. 

The Kitaev chain model is a fundamental topological superconducting model in condensed matter physics and quantum computation, known for its role in the study of topological phases and predicting the existence of Majorana Zero Modes (MZMs) at the ends of the chain \cite{AKitaev_2001}. These MZMs are quasiparticles that are their antiparticles, and they emerge as zero-energy edge states in the topologically non-trivial phase of the model. These modes are topologically protected, meaning they are robust against local perturbations such as disorder, making them ideal candidates for encoding quantum information. MZMs in the Kitaev chain exhibit non-Abelian statistics, a key feature for topological quantum computation \cite{PhysRevB.85.035110}. In this approach to quantum computing, information is stored in the global state of the system, making it resistant to local errors. Braiding MZMs, which can manipulate quantum information in a topologically protected manner, is a foundational concept in developing fault-tolerant quantum computers  \cite{Sarma2015-av,Flensberg2021-ae}. The concepts introduced by the 1D Kitaev chain have inspired significant experimental efforts to realize topological superconductivity and MZMs. For instance, systems such as semiconductor-superconductor nanowires \cite{PhysRevB.103.045428,Stanescu_2013,PhysRevB.107.035440}, where strong spin-orbit coupling is present, have been proposed and experimentally investigated as platforms that could realize the Kitaev chain model. Observations of zero-bias conductance peaks \cite{PhysRevB.98.155314,PhysRevB.101.024506,PhysRevLett.129.167702,PhysRevB.96.184520,PhysRevB.106.094504,PhysRevB.101.024506} in these systems are considered possible experimental signatures of MZMs. Over the past two decades, a well-studied system proposing to host MZMs has gathered significant interest in the condensed matter community.

In this work, we have investigated the Kitaev chain on a square lattice with next-nearest-neighbor (NNN) hopping and pairing. This model is equivalent to the two-dimensional extension of the Kitaev chain in the absence of the NNN coupling term. Our work brings several interesting phenomena together. The structure of the paper is as follows: To begin with, we introduce the model and review it under the absence of the NNN-coupling strength (refer Sec. \ref{Model}). Later, in Sec. \ref{sec3}, we study the system under various boundary conditions. In Sec. \ref{Defects}, we examine how varying $\eta$ leads to different topological phase transitions, with gapless BTPs characterizing the system's topology. We analyze the system under Periodic Open Boundary Conditions (POBC) in Sec. \ref{POBC} to further investigate the edge states. While the edge spectrum under POBC confirms the presence of edge states, their precise location can only be determined under Open Boundary Conditions (OBC). Therefore, we also examine the system under OBC in Sec. \ref{OBC}, confirming both the presence and pinpointing the edge states' location.  

\section{Model}
\label{Model}

\begin{figure}
\centering\includegraphics[width=1\columnwidth]{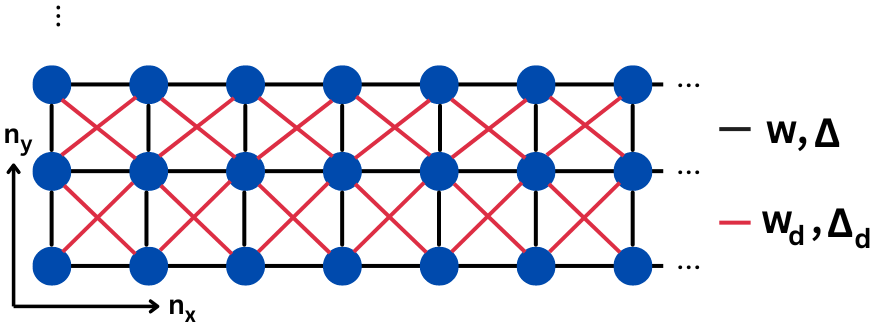}
\caption{Schematic diagram of the Kitaev chains on a square lattice with inter-chain NNN hopping and pairing.}
\label{NNN}
\end{figure}

We now discuss the $p$-wave Kitaev chains on a square lattice. This system is our prototype model, which will be studied to identify the MZMs and qMMs and investigate how to distinguish these two modes. In this section, we first describe the model in detail and then discuss different boundary conditions. In one case, we consider periodic boundary conditions (PBCs) along both directions. We refer to this boundary condition as the system in toroidal geometry. For this case, we can study the bulk properties of the energy bands. Since we are interested in investigating the edge states of the system, we also consider the system with open boundary conditions (OBC). Here, besides OBC along both directions, we consider PBC along one direction and OBC along the other. This partial OBC (POBC) case can further be divided into two sub-cases depending on the direction of the OBC and PBC.

\subsection{The $p$-wave Kitaev chains on a square lattice}
  
In this model, we consider intra-chain and inter-chain NN hopping and pairing. Without losing generality, we set the NN hopping and pairing strengths equal for $x-$ and $y-$directions \cite{2DKitaev,zhang2019}. Moreover, we have introduced inter-chain NNN hopping and pairing. A schematic diagram of this model is presented in Fig. \refb{NNN}. Here, we consider the ``$p_x + p_y$'' type of pairing. The Hamiltonian of the model is given as: 
\begin{widetext}
\begin{subequations}
\label{H}
\be
H = H_{\rm 2D-Kitaev} + H_{\rm NNN},
\ee
$H_{\rm 2D-Kitaev}$ represents the Kitaev chain Hamiltonian on the square lattice, whereas $H_{\rm NNN}$ is the Hamiltonian corresponding to the NNN hopping and pairing. The Hamiltonian $H_{\rm 2D-Kitaev}$ can be expressed as
\be
H_{\rm 2D-Kitaev} = H^{\rm onsite} + H^{\rm hopping} + H^{\rm pairing},
\ee
where
\be
\begin{split}
H^{\rm onsite} &= \mu \sum_{n_x,n_y=1}^{N}  c_{n_x,n_y}^\dag c_{n_x, n_y}\\
H^{\rm hopping} &= \sum_{n_x,n_y=1}^{N-1}  w \Big( c_{n_x,n_y}^\dag c_{n_x+1,n_y} + c_{n_x,n_y}^\dag c_{n_x,n_y+1}\Big) + {\rm H.c.}\\
H^{\rm pairing} &= \sum_{n_x,n_y=1}^{N-1} \Delta  \Big( c_{n_x,n_y}^\dag c_{n_x+1,n_y}^\dag + c_{n_x,n_y}^\dag c_{n_x,n_y+1}^\dag \Big) + {\rm H.c.},
\end{split}
\ee
where the operators $c_{n_x,n_y}^\dag (c_{n_x,n_y})$ creates (annihilates) a spin-less fermion at ($n_x,n_y$) site, $\mu$ is the chemical potential, $w$ is the nearest-neighbor (NN) hopping strength, and $\Delta$ is the NN pairing strength. Here, we have set the number lattice sites along $\widehat{x}$ and $\widehat{y}$ directions equal, i.e., $N_x = N_y = N$. Moreover, we have considered the strength of hopping and pairing along $\widehat{x}$ and $\widehat{y}$ directions equal. We write the NNN part of the Hamiltonian $H_{\rm NNN}$ also as a sum of its hopping and pairing terms as:
\be
H_{\rm NNN} = H_{\rm NNN}^{\rm hopping} + H_{\rm NNN}^{\rm pairing},
\ee
where
\be
\begin{split}
H_{\rm NNN}^{\rm hopping} &= \sum_{n_x,n_y=1}^{N-1} w_d \Big( c_{n_x,n_y}^\dag c_{n_x+1,n_y+1} + c_{n_x+1,n_y}^\dag c_{n_x,n_y+1}\Big)  + {\rm H.c.}\\
H_{\rm NNN}^{\rm pairing} &= \sum_{n_x,n_y=1}^{N-1} \Delta_d \Big( c_{n_x,n_y}^\dag c_{n_x+1,n_y+1}^\dag + c_{n_x+1,n_y}^\dag c_{n_x,n_y+1}^\dag \Big) + {\rm H.c.}
\end{split}
\ee
\end{subequations}
The conventional way to deal with the pairing term is to work in the space-doubled Bogoliubov-de Gennes (BdG) basis \cite{Kitaev}. In this basis, the system's Hamiltonian is written as:
\begin{subequations} 
\be
H =  \Psi^\dag \mathcal{H}_{\rm BdG} \Psi, 
\ee
where
\be
\Psi = [c_{1,1},\dots,c_{1,N},c_{2,1},\dots,c_{N,N}, c_{1,1}^\dagger,\dots,c_{1,N}^\dagger,c_{2,1}^\dagger,\dots,c_{N,N}^\dagger]^{\rm T},
\ee
\end{subequations}
and ${\rm T}$ denotes transpose of the vector. The Hamiltonian kernel $\mathcal{H}_{\rm BdG}$ is a $2M \times 2M$ dimensional matrix, where $M = N^2$. In this work, we set the NNN hopping and pairing strengths equal, i.e., $w_d = \Delta_d = \eta$.
\end{widetext}

\subsection{Boundary conditions}

\subsubsection{Toroidal geometry}
 
\begin{figure}[t]
\centering\includegraphics[width=1\columnwidth]{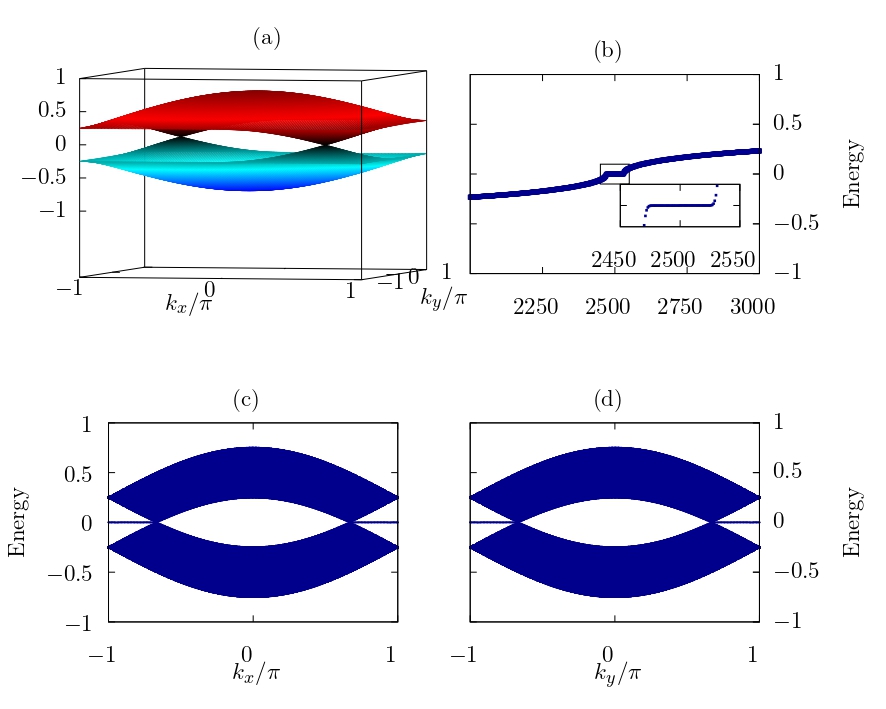}
\caption{Panel (a) displays the bulk band spectrum of the Hamiltonian on Eq. \refb{h_k} under PBC,  and (b) displays the energy eigenvalue spectrum of the $50 \times 50$ square lattice Hamiltonian on Eq. \refb{H} under OBC. The inset panel confirms the presence of multiple zero-energy eigenvalues or MZMs. (c) Displays the energy eigenvalue spectrum of the Hamiltonian under POBC, where PBC is considered along $x$-direction and OBC along the $y$-direction with $N_y = 200$ lattice sites. (d) Displays the energy eigenvalue spectrum of the Hamiltonian under POBC, PBC along $y$-direction, and OBC along $x$-direction with $N_x = 200$ lattice sites.} 
\label{2D Kitaev}
\end{figure}

The bulk properties of the model are explored by imposing periodic boundary conditions along both directions. We apply the Fourier transformation \cite{NSSH}
\begin{subequations}
\label{h_k}
\be 
c_{n_x,n_y}^\dag = \frac{1}{N^2} \sum_{k_x,k_y } e^{-i(k_x n_x + k_y n_y)} c_{k_x,k_y}^\dag,
\ee
and get the Hamiltonian in ${\bs k}$-space as  
\be
H = \sum_{\bs k>0} \Psi_{\bs k}^\dag \mathcal{H}_{\bs k} \Psi_{\bs k},
\ee 
where $\Psi_{\bs k} = [c_{\bs k}~~ c_{-{\bs k}}^\dag]^T$ is the Nambu spinor corresponding to the BdG Hamiltonian in the ${\bs k}$-space and the Hamiltonian kernel $\mathcal{H}_{\bs k} = {\bs h}_{\bs k} \cdot {\bs \sigma}$. Here, ${\bs \sigma} = (\sigma_x, \sigma_y, \sigma_z)$ are the pseudo-spin Pauli matrices. The components of ${\bs h}_{\bs k}$ are
\be 
\begin{split}
h_x({\bs k}) &= 0, \\
h_y({\bs k}) &= \Delta (\sin k_x + \sin k_y ) + 2 \eta \cos k_x \sin k_y,  \\
h_z({\bs k}) &= \mu + w (\cos k_x + \cos k_y) + 2 \eta \cos k_x \cos k_y.
\end{split} 
\ee
\end{subequations}
The intrinsic properties of the system, like bulk-energy bands, topological invariants, and electronic states of the bands, are studied using the bulk Hamiltonian given in Eq. \eqref{h_k}. The band spectrum of the bulk Hamiltonian is shown in Fig. \ref{2D Kitaev}(a) \cite{PhysRevB.83.125109}. The boundary properties, like the edge, corner, and surface states, can be seen only under OBC conditions. Due to the complexity of the Hamiltonian matrix under OBC, we initially begin by analyzing the system using cylinder geometry or the partial open boundary conditions (POBC), in which, at each instant, one direction follows OBC and the other follows PBC.

\begin{widetext}
\subsubsection{Cylindrical geometry}
For the POBC case, first, we consider OBC along $x$-direction and PBC along $y$-direction. Corresponding Hamiltonian becomes:
\be  
H(k_x) =  \Bigg\{ \Big( [\mu + w \cos k_x]\sigma_z - \Delta \sin k_x \sigma_y \Big) \otimes \mathbbm{1}_{N_y} + \Big( [w + \eta \cos k_x] \sigma_z  + [\Delta + \eta \cos k_x] \sigma_y  \Big) \sum_{n_y} c_{n_y}^\dag c_{{n_y}+1}^\dag   +H.c. \Bigg\}. 
\label{h1}
\ee
Similarly, we consider PBC along $y$-direction and OBC along $x$-direction. For this case, the Hamiltonian becomes:
\be 
H(k_y) =  \Bigg\{ \Big( [\mu + w \cos(k_y)]\sigma_z - \Delta \sin(k_y) \sigma_y \Big) \otimes \mathbbm{1}_{N_x}  + \Big( [w + \eta \cos(k_y)] \sigma_z  +[ \Delta +\eta \cos(k_y)] \sigma_y \Big) \sum_{n_x} c_{n_x}^\dag c_{{n_x}+1}^\dag + H.c. \Bigg\}, 
\label{h2}
\ee
\end{widetext}
where $\mathbbm{1}_{N_x} (\mathbbm{1}_{N_y})$ denotes the $ N_{x (y)} \times N_{x (y)}$ identity matrix. Our primary focus is on the role of the coupling strength $\eta$. We probe how the variation of $\eta$ induces new topological phases, thereby enriching the system's topological characteristics. 

\section{Results and Discussions}
\label{sec3}

We begin our analysis of the system considering the absence of the NNN hopping and pairing terms in the Hamiltonian by setting $\eta = 0$. This study is crucial to understanding the influence of the NNN terms on the system. Throughout this work, we set the parameters at $\mu = 1.0,\, w = 1.0$, and $\Delta = -1.0$. Figure \ref{2D Kitaev}(a) shows that the system is gapless with two band touching points (BTPs). In Fig. \ref{2D Kitaev}(b), we consider the system with $50 \times 50$ square lattice under the OBC along both directions. The inset plot reveals the presence of multiple degenerate edge states. We further explore the edge or boundary properties of the system by analyzing the Hamiltonian under POBC (refer to Eqs. \eqref{h1} and \eqref{h2}) in Figs. \ref{2D Kitaev}(c)-(d). We now investigate the topological and localization properties of the system that emerge due to introducing new coupling terms. 

\subsection{Topological Defects}
\label{Defects}

\begin{figure*}
\includegraphics[width=1\textwidth]{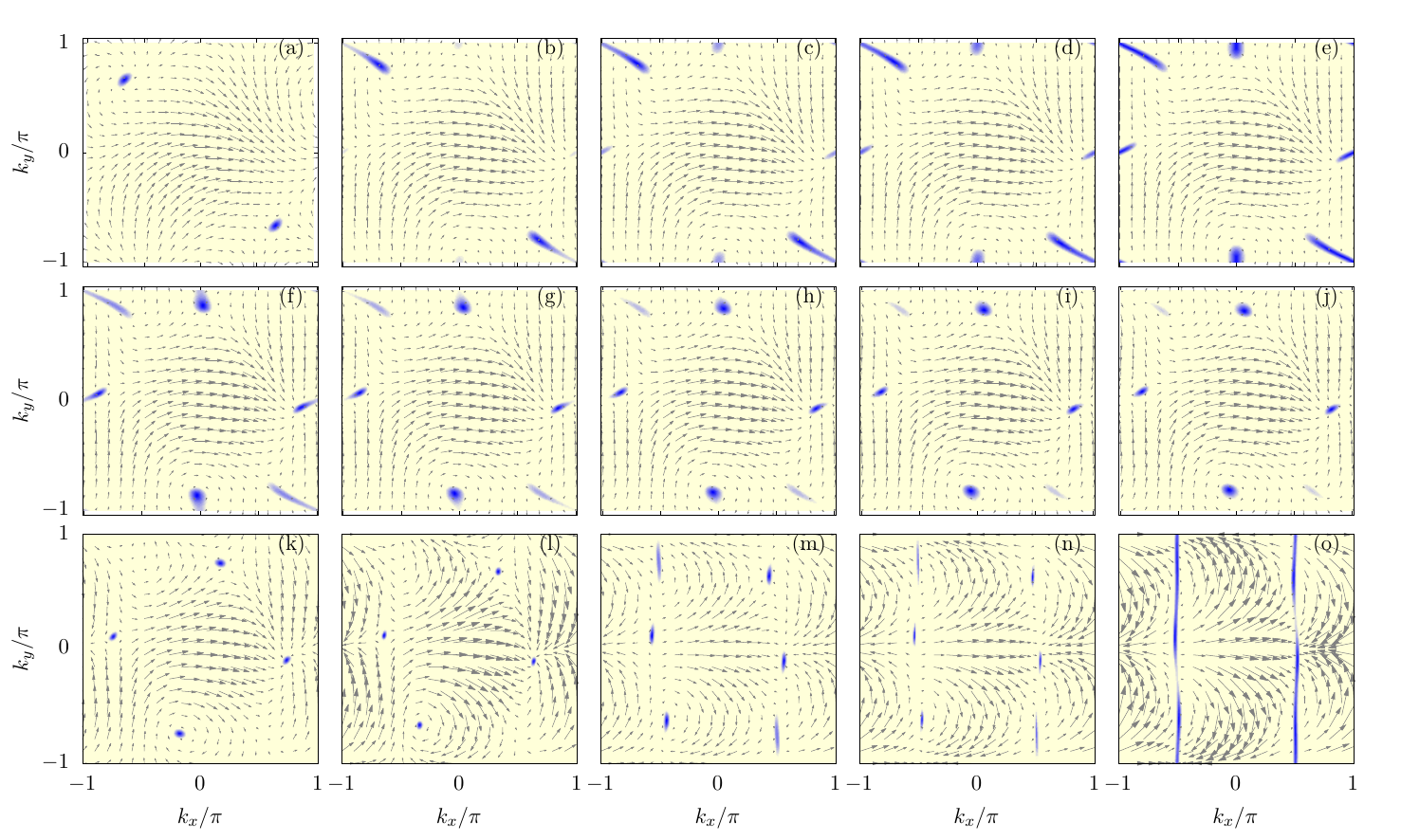}
\caption{This figure displays the Berry curvature of the Hamiltonian in Eq.  \refb{H_kvec}, which is shown by calculating the direction (denoted by the vectors) and magnitude (denoted by the color) against the components of the Hamiltonian. Panels \Blue{(a-e)} shows the shifting of the singularity points by varying $\eta = (0, 0.4, 0.43, 0.45, 0.5)$. Panels \Blue{(f-j)} is for $\eta = (0.6, 0.63, 0.65, 0.68, 0.7)$. Panels \Blue{(k-o)} is for $\eta = (1, 2, 5, 10, 20)$, which shows how the singularity points are aligned along a single line as the value of $\eta$ is comparably very high.}
\label{pbc_vec}
\end{figure*}

\begin{figure*}
\includegraphics[width=1\textwidth]{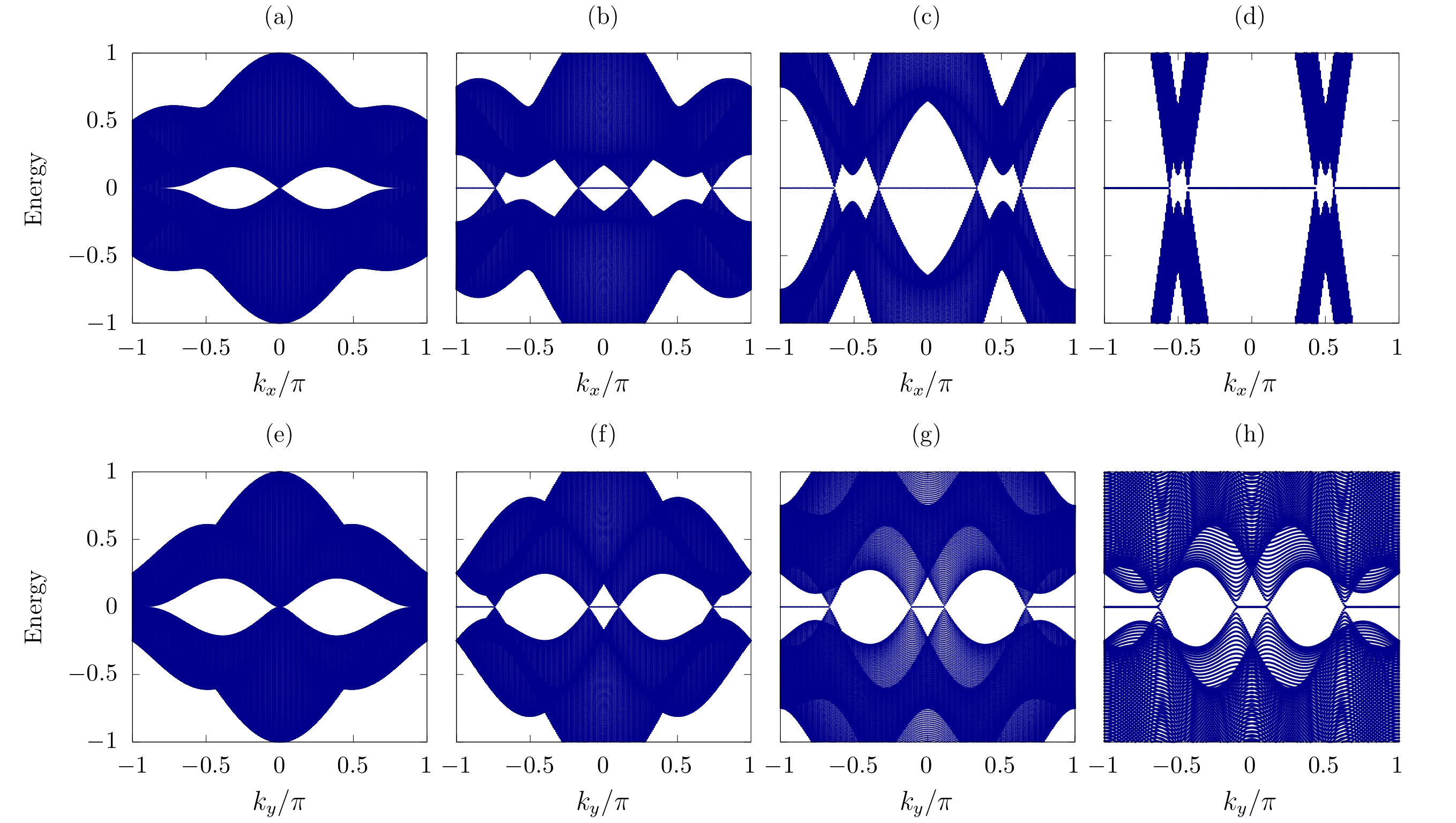}
\caption{The edge spectrum for the Hamiltonian in Eq. \refb{maj_ham1} is shown in plots (a-d). The same spectrum for the Hamiltonian given in Eq.  \refb{maj_ham2} is shown in panels (e)-(h). By viewing the system in cylinder geometry, we can confirm the presence of multiple MZMs in the system for different cases. The NNN coupling strength $\eta= (0.5,1,2,5)$ for (a-d) and (e-h), respectively. Here, we set $N_x=N_y=200$.}
\label{pobc}
\end{figure*}

The gapless BTPs in condensed-matter systems are identified as singularities in the Brillouin zone, elucidating the system's Berry curvature. The underlying symmetry-induced degeneracy leads to the Dirac or Weyl points. Then, by examining the components of the Hamiltonian as a vector field, we observe vortices in the field, and those lead to two-dimensional topological defects like the Dirac cones or Weyl nodes \cite{Hasan2021-qz,Jia2016-qo}. They do not carry a topological charge because they are composed of two overlapping Weyl points of opposite chirality, canceling each other's topological charge. On the other hand, a Weyl point can be seen as half of a Dirac point when the system's time-reversal (TRS) or inversion symmetry is broken. They carry a topological charge, known as the Chern number, which can be positive or negative depending on the chirality of the Weyl point. This charge represents the ``monopole'' nature of the Weyl points in momentum space. They are responsible for the emergence of flux-carrying vortices (or monopoles) in the system \cite{Funfhaus_2022}. Our model exhibits multiple gapless BTPs responsible for the system's two-fold degenerate zero-energy points. These degenerate points are localized in the vortex core, leading to topological defects in the 2D BZ. We mark these defects in Fig. \refb{pbc_vec} by analyzing the Hamiltonian components and locating how the BTPs lead to topological defects in the system. Panel \ref{pbc_vec}\Blue{(a)} displays the two topological defects (denoted by blue color points) corresponding to two BTPs present in the Kitaev chain on a square lattice [refer Eq. \eqref{h_k} and Fig. \ref{2D Kitaev}]. These two topological defects are robust against disorder and cannot be removed unless the band gap opens and closes due to variation in system parameters \cite{zhang2019}. To further reveal similar properties in our model, we see the components of the Hamiltonian and observe the emergence of new singularities (or the monopoles) near the vicinity of the BTPs ($k_{x_0},k_{y_0}$) in the bulk Hamiltonian,
\be
 \begin{split} 
 & \Delta [ \sin(k_{x_0}) + \sin(k_{y_0}) ] + 2\eta \cos (k_{x_0}) \sin (k_{y_0}) =0 \\&
  \mu + w [ \cos(k_{x_0}) + \cos(k_{y_0})] + 2\eta \cos (k_{x_0}) \cos (k_{y_0})] =0 
  \end{split} 
  \label{H_kvec}
\ee
The above equation suggests that, for $(k_{x_0}, k_{y_0}) = (0, \pm \pi)$, we have $\eta = \frac{\mu}{2}$, whereas for $(k_{x_0}, k_{y_0}) = (\pm \pi, \mp \pi)$, we have $\eta = \frac{2w - \mu}{2}$. For our choice of system parameters, when $\eta = 0.5$, both conditions are met. This leads to the emergence of a new bulk topological phase, which corresponds to the appearance of new topological defects (refer to Fig.  \ref{pbc_vec}\Blue{(e)}). We analyze the model by varying $\eta$ to highlight the changes in the defects. Figures \ref{pbc_vec}(a)-(e) illustrate how the topological defects shift and rise in numbers as $\eta$ increases from $0$ to $0.5$. Here, $\eta = 0.5$ is the critical point where the model transitions from a topological gapless state with two BTPs to a topological gapless state with four BTPs. Further increment of the parameter $\eta$, as shown in Figs. \ref{pbc_vec}(f)-(j) reveals how the defects move towards the center of the BZ as $\eta$ varies from $0.6$ to $0.7$. The dark blue spots represent the BTPs, while the light blue spots indicate that the band gaps are approaching zero. The topology of these states will be analyzed in subsequent sections.

Figures \ref{pbc_vec}(k)-(o) demonstrate how the four defects align along a straight line, independent of the momentum $k_y$ when $\eta$ becomes sufficiently large. In this case, the point defects evolve into a nodal line. Regarding topological charge, we observe that the defects always appear in even numbers with opposite chirality, as indicated by the direction of the arrows in Fig. \ref{pbc_vec}. These points can act as sources (positive chirality) or sinks (negative chirality) of the Berry curvature  \cite{WeylSSH}. This result confirms that all the BTPs are Bogoliubov-Weyl nodes (BWNs) with zero Chern number, indicating that the model is a weak topological superconductor (TSC) \cite{Aperiodic_weakTSC,weakTSC1,weakTSC2}. This effect is purely topological, stemming from the band structure, as the bulk Fermi surface vanishes at the BWNs. The Dirac point (a pair of Weyl points with opposite chirality) with four-fold degeneracy is generated when two doubly degenerate energy bands cross. Throughout the variation of $\eta$, the number of BTPs remains even, always occurring in pairs of BWNs, suggesting that the system exhibits characteristics of a Dirac semimetal (DSM) \cite{2D_Dirac,CA_DSM}.

\subsection{Dirac Semimetal on a TSC Phase}
\label{POBC}

In the previous section, we have shown the presence of BWNs in the model at different $(k_x, k_y)$ values depending on the parameter $\eta$. We now shift our focus to the model's boundary properties to reveal the topological nature of BWNs and emphasize the characteristics of DSMs more. Unlike the WSMs, the DSMs do not have surface Fermi arcs \cite{Yan2017-cp, 3DW_DSM}. This happens due to an even number of pairs of BWN, which makes the net topological charge zero. The literature suggests that the topological superconducting phase arises when a DSM and a TSC come into contact \cite{TSC_in_DSM,TSC_in_DSM2}. One of the most intrusive possibilities is the emergence of MZMs at the edges or surface of the contact. In order to reveal the existence of MZMs, we present the edge spectrum for two distinct conditions utilizing the Majorana basis. These bases give us a clear demonstration of the presence of multiple MZMs and different topological phases obtained in terms of the coupling strength $\eta$. For the OBC case, we substitute the Dirac operators $c_{n_x,n_y}$ by the Majorana operators $(a_{n_x,n_y}, b_{n_x,n_y})$ in the Hamiltonian, where these two sets of operators are related by
\begin{widetext}
\be
\begin{split} 
c_{n_x,n_y}^\dag &= a_{n_x,n_y} - i b_{n_x,n_y}\\
c_{n_x,n_y} &= a_{n_x,n_y} + i b_{n_x,n_y}. 
\end{split}
\ee
In terms of the Majorana operators, the Hamiltonian of the system, given in Eq. \eqref{H}, becomes $H = H_{\rm{2D-Kitaev}} + H_{\rm{NNN}}$, where
\be
\begin{split}
H_{\rm{2D-Kitaev}} &= \frac{i}{4} \sum_{n_x,n_y}  \Big\{ 2\mu a_{n_x,n_y} b_{n_x,n_y}+ J( a_{n_x+1,n_y} b_{n_x,n_y} + a_{n_x,n_y+1} b_{n_x,n_y} )  + Q (a_{n_x,n_y} b_{n_x+1,n_y}  + a_{n_x,n_y} b_{n_x,n_y+1}) + {\rm H.c.} \Big\}\\
H_{\rm NNN} & = 2\eta\sum_{n_x,n_y=1}^{N-1} \Big( a_{n_x+1,n_y+1} b_{n_x,n_y} + a_{n_x,n_y+1} b_{n_x+1,n_y}\Big) + {\rm H.c.}
\label{maj_ham}
\end{split}
\ee
Here, we define the parameters $J = w + \Delta$ and $Q = w - \Delta$.

\begin{figure*}[t]
\includegraphics[width=1\textwidth]{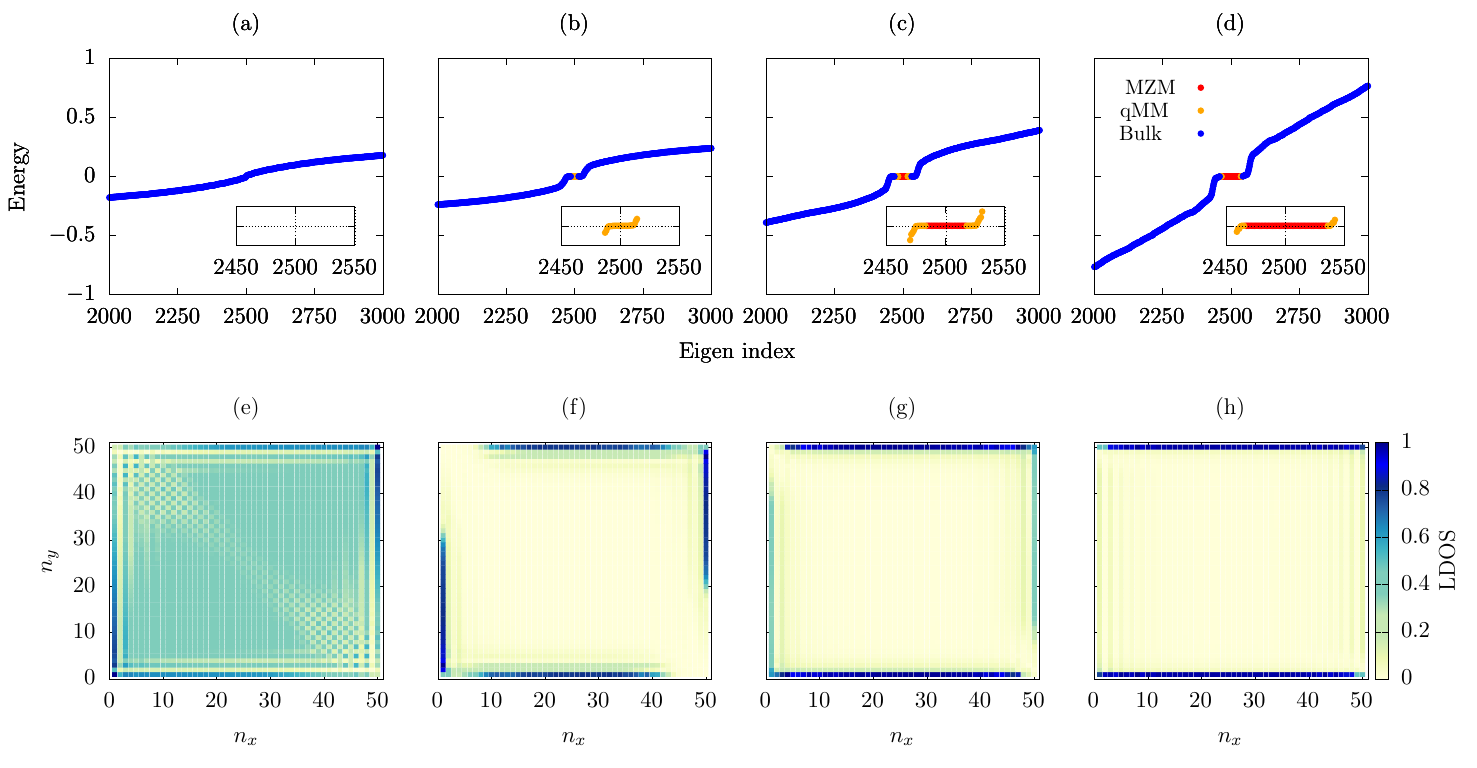}
\caption{Energy eigenvalues of a system of $50 \times 50$ square lattice, whose Hamiltonian is given in Eq.  \refb{maj_ham}, are shown in panels (a)-(d). The inset plot confirms the presence of multiple gapped-zero energy states in the system. Like them, low-energy or nearly-zero-energy states are also visible in the inset plot. The energy range of the inset plot is $|E|=10^{-5}$. The localization of the zero energy states is confirmed in panels (e)-(h) by presenting the site-dependent normalized linear density of states (LDOS), where $n_x$ and $n_y$ are the lattice site indices. The dark blue spots reveal the presence of MZMs, and the light blue reveals the presence of qMMs in the systems. The light yellow colored region shows the delocalized or paired fermionic states. In this plot, we have varied $\eta=$(0.5,1,2,5) along (a-d) and (e-h) respectively.}
\label{obc_ldos}
\end{figure*}

We diagonalize the system under cylinder geometry (or POBC) to get the edge spectrum that shows MZMs. The POBC allows the Hamiltonian to be written in two different cases, similar to the previous section. Case I: The OBC is considered along the $x$-direction, and PBC along the $y$-direction, and the corresponding Hamiltonian becomes:
\be 
H = \frac{i}{4} \sum_{n_x} \sum_{k_y}  \Big\{  (2\mu +J e^{ik_y} + Q e^{-ik_y} ) a_{n_x,k_y} b_{n_x,-k_y} + (J +2\eta e^{ik_y} ) a_{{n_x}+1,k_y} b_{{n_x},-k_y}  + (Q +2\eta e^{ik_y} ) a_{{n_x},k_y} b_{{n_x}+1,-k_y} + {\rm H.c.} \Big\} 
\label{maj_ham1} 
\ee
Case II: The OBC is considered along $y$-direction and PBC along $x$-direction, and the Hamiltonian for this case is:
\be  
H = \frac{i}{4} \sum_{k_x} \sum_{n_y} \Big\{(2\mu +J e^{ik_x} +Q e^{-ik_x} ) a_{k_x,{n_y}} b_{-k_x,{n_y}} + [J +2\eta \cos(k_x) ] a_{k_x,{n_y}+1} b_{-k_x,{n_y}} + Q a_{k_x,{n_y}} b_{-k_x,{n_y}+1} + {\rm H.c.} \Big\} 
\label{maj_ham2} 
\ee
\end{widetext}
We show the edge spectrum corresponding to Case I in the first row of Fig. \refb{pobc}. The second row shows the edge spectrum for Case II. From the previous sections and as illustrated in Figs. \ref{2D Kitaev}(c)-(d), it is evident that in the absence of the parameter $\eta$, the system remains in a gapless topological state for the chosen system parameters. So, by systematically varying the parameter $\eta$, we show that the system undergoes a phase transition through a gap-closing and opening. At $\eta=0.5$, Figs. \ref{pobc}(a, e), the system is in a trivial state with no edge states. This is the critical point (as discussed in Sec. \ref{Defects}), where the system makes a transition from one topological phase with {\it two} BWNs to another topological phase with {\it four} BWNs. A careful analysis of the energy eigenstates corresponding to these zero-energy states reveals that these states are indeed MZMs. This fact is the direct consequence of the characteristics of a DSM on a TSC system  \cite{TSC_in_DSM,TSC_in_DSM2}. Unlike the WSM, this system does not host Fermi arc surface states. The absence of Fermi arc states and the presence of MZMs confirm that our model is a DSM.

Following the results obtained from the previous section, upon increasing $\eta$, the BTPs change their location, but their numbers do not change. Moreover, as $\eta$ increases, we see an increment in the number of zero energy states when PBC is taken along $x$-direction, as presented in Figs. \ref{pobc}(c)-(d). This result implies that some degenerate edge states are more localized along the ends of vertical chains. The purpose of looking at the system under POBC was to show the presence of MZMs and the absence of Fermi arcs. 

\begin{figure*}
\includegraphics[width=1\textwidth]{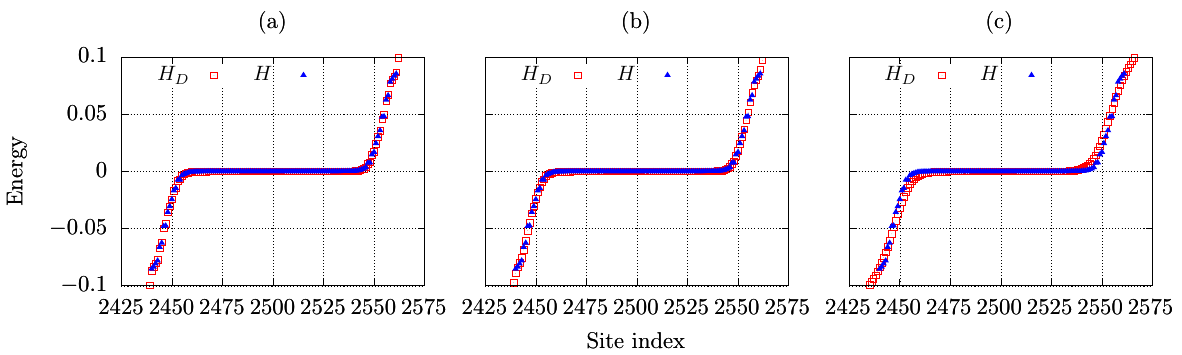}
\caption{The figure displays the robustness of the edge states against disordered potential given in the system. The Hamiltonian under the OBC with $50 \times 50$ lattice sites is investigated by introducing the disorder parameter $\zeta$. The red squares denote the spectrum of the disordered Hamiltonian $H_D$, whereas violet triangles represent the spectrum of the unperturbed Hamiltonian $H$. The disorder strength is modified in each panel as (a)$\zeta= 0.01$, (b)$\zeta= 0.1$, and (c)$\zeta= 0.5$.}
\label{disorder_obc}
\end{figure*}

\subsection{True Topological MZMs and qMMs}
\label{OBC}

In this subsection, we investigate the nature of the localization in the system by studying it under full OBC. For this purpose, we compute the energies under OBC for a square lattice with the number of lattice sites along $x$-direction $N_x = 50$ and $y$-direction $N_y = 50$. In the Majorana basis, \[\Psi = [a_{1,1} \,\dots\, a_{N,1}\, a_{1,2}\,\dots\,a_{N,N}\, b_{1,1}\,...\,b_{N,1}\, b_{1,2}\,\dots\,b_{N,N} ]^T,\] the dimension of the Hamiltonian kernel will be $2M \times 2M$, where $M = N_x \times N_y = 2500$. Figures \ref{obc_ldos}(a)-(d) display the energy spectrum under OBC for different values of the parameter $\eta$. The inset of the figures reveals the location of different edge states with different color points. The blue colored points represent the bulk part of the Hamiltonian. Depending upon the accuracy $|E| \leq 10^{-5}$ and the localization, we have categorized the edge states into two types: (i) true topological MZM  with exact zero energy and (ii) qMMs with nearly zero energy Andreev bound states \cite{qMM1,qMM2}. Both the edge states can be differentiated from the bulk. The red points correspond to the {\it true} topological MZMs, and the orange points denote the qMMs. We also compute the normalized site-resolved linear density of states (LDOS), given by  \[N_i(E) = \sum_{n} |\phi_n(i)|^2 \delta(E-E_n),\] 
where $n$ is the eigenvalue index with eigenvalue $E_n$ and $\phi_n(i)$ is the eigenstate of each site-index $i$ \cite{Arijit_subha}. For $\eta=0.5$, the inset of Fig. \ref{obc_ldos}(a) confirms that it is a trivial phase with zero MZMs. It is evident from the LDOS results, given in Fig. \ref{obc_ldos}(e), that the system is completely delocalized. The zero energy states appear as the value of $\eta$ increases. Furthermore, when $\eta$ increases, the inset of Fig. \ref{obc_ldos}(b) reveals the presence of non-degenerate eigenvalues. 

The LDOS results show that the qMMs are also localized and, at specific conditions, mimic the MZMs \cite{qMM1,qMM2}. Figure \ref{obc_ldos}(f) shows the result for $\eta = 1.0$. The system is in a topological state with multiple MZMs and qMMs. The dark-blue squares denote the completely localized states representing MZMs, whereas the light-blue squares indicate nearly localized states representing qMMs. We further increase the parameter $\eta$; the corresponding results are presented in Figs. \ref{obc_ldos}(g)-(h). These figures confirm that the edge states are localized along one direction, as observed under POBC in Figs. \ref{pobc}(c) and (g), and also in (d) and (h). Figure \ref{obc_ldos}(h) confirms that the MZMs are localized at the ends of the vertical chains. The Hamiltonian under OBC confirms the signature of Majorana modes on a DSM in the TSC phase.

\subsection{Robustness of Topological Edge states}
\label{disorder}

We know that, unlike the MZMs, the qMMs are not topologically protected. Therefore, we can distinguish these two modes by validating the robustness of the topological protection in the presence of disorder \cite{2DKitaev}. For this purpose, we introduce a random perturbation to the system parameters of the form 
\be 
\begin{split} 
\mu & \rightarrow \mu_D = \mu (1 + \zeta_1) \\
 w & \rightarrow w_D = w(1 + \zeta_2)  \\ 
 \Delta & \rightarrow \Delta_D = \Delta(1 + \zeta_3), 
\end{split} 
\ee
where $\zeta_i$'s with $i = 1, 2, 3$ are uniformly distributed random numbers in the interval $[-\zeta, + \zeta]$. Note that the disorder parameter $\zeta$ is chosen to be real to preserve the TRS in the system. Consequently, the DSM nature of the system remains unchanged. The investigation of the energy spectrum of the disordered Hamiltonian $H_D = H(\mu_D, w_D, \Delta_D,\eta)$ is shown in Fig. \ref{disorder_obc}. In this figure, all the results represent an ensemble average of $1000$ realizations. Here, we see the robustness of MZMs against the disordered potential. The red squares denote the spectrum of the disordered Hamiltonian $H_D$, whereas violet triangles represent the spectrum of the unperturbed Hamiltonian $H$. We observe that the edge states do not vary significantly for the disorder strength $\zeta = 0.01$. However, Fig. \ref{disorder_obc}(a) shows that the violet triangles and red squares do coincide, which indicates that the disorder affects the bulk energy with $|E| > 0.05$. Since this amount of disorder is not sufficient to affect the edge states, we increase the disorder strength to $\zeta = 0.1$. Figure \ref{disorder_obc}(b) presents the result corresponding to this stronger disorder. Here, we see that, due to the disorder, the energies of the unperturbed and perturbed systems do not coincide in the range $|E|> 10^{-3}$. This observation reveals the lack of topological protection in the qMMs but not in the MZMs. Consequently, this distinguishes the true topological MZMs and qMMs concerning topological protection. Figure \ref{disorder_obc}(c) suggests that further increase in disorder with $\zeta = 0.5$, more number of qMMs have moved from edge to bulk part. Even for this strong disorder, the MZMs are still localized at the edges, preserving their topological properties. This analysis is carried out for NNN hopping and pairing strengths $\eta = 2.0$, where the corresponding LDOS results depict the presence of both MZMs and qMMs in the system as shown in Fig. \ref{obc_ldos}(g). From this study of the system in the presence of disorder, we infer that the edge states corresponding to the qMMs mix with the bulk for sufficiently strong disorder. In contrast, the MZMs still preserve their topological nature.

\section{Summary and conclusion}

In this work, we have investigated the Kitaev chain on a square lattice with NNN hopping and pairing. Here, we focus on how the NNN terms enrich the system's topology with the advent of new topological phases. The analysis begins with studying the system without the NNN terms, which gives a crucial understanding of the role of these terms in the system. We investigate the system under three different boundary conditions: (i) PBC (toroidal geometry), (ii) POBC (cylindrical geometry), and (iii) OBC (flat geometry). Initially, under the PBC, we see that bulk spectrum hosts gapless topological states with two BTPs in the 2D BZ \cite{2DKitaev}. They depict the presence of BWNs for different strengths of the NNN terms \cite{WeylSSH}. Here, the system transitions from a topological state with two BWNs to another topological state with four BWNs. We show that the BWNs always occur in even pairs, revealing the characteristics of Dirac semimetal \cite{2D_Dirac}. This result is highlighted by analyzing the system under the POBC, which shows the presence of Majorana state as a direct consequence of a DSM in a TSC phase. Also, by probing the system under the OBC, we have elucidated the localization of topological edge states obtained in the system. Depending upon their accuracy to be zero energy and localization, we have distinguished the edge states into two modes: (i) true topological MZMs and (ii) qMMs.

Further, we have confirmed the nature of the topological protection of the edge states by introducing random disorder in the system. This study confirms that the true topological edge states are robust against the disorder \cite{zhang2019}. Furthermore, our study confirms that the model hosts BWNs with a zero Chern number, indicating the weak topological superconducting nature of the system. Considering the potential applications of the topological superconducting materials, we have investigated a model with DSM characteristics and a weak topological superconductor hosting multiple topological edge states. 

Finally, we conclude with a short note on the probable experimental setups that can realize the Kitaev chain on a square lattice system. Ongoing research activities explore various Dirac semimetals and their interactions with superconductors and other materials to realize and understand the properties of qMMs \cite{TSC_in_DSM,TSC_in_DSM2}. This decade has opened up new techniques to realize topological superconductivity in minimal Kitaev chain devices. Notably, a proposal has been made to extend the Kitaev chain to include longer-range interactions \cite{Kitaev_exp}. Topological electronics or ``topo-electronics" has garnered significant interest within the quantum transport community, offering a compelling approach to realizing topological materials. Recently, a topological integrated circuit (TIC) implementation has observed the spatial profile of a topological edge state in a 1D Kitaev model at a resonant frequency of $13$ GHz \cite{top_circuits}. This circuit representation can be extended to the 2D version.

\acknowledgments
JNB acknowledges financial support from DST-SERB, India, through a MATRICS grant MTR/2022/000691.

\end{document}